\documentclass[journal]{IEEEtran}

\usepackage{amsmath,amssymb,amsfonts}
\usepackage{bm}
\usepackage{cite}
\usepackage{graphicx}
\usepackage{tikz}
\usetikzlibrary{arrows.meta,calc,positioning,shapes.geometric}
\usepackage{booktabs}
\usepackage{multirow}
\usepackage{array}
\usepackage{threeparttable}
\usepackage{url}

\usepackage{algorithm}
\usepackage{algorithmic}

\usepackage[caption=false,font=footnotesize]{subfig}

\makeatletter
\def\@IEEEtablecaptionsepspace{\vskip -0.35\baselineskip\relax}
\makeatother

\usepackage[hidelinks]{hyperref}


\makeatletter
\renewcommand\@makefntext[1]{\noindent\@makefnmark{}#1}
\makeatother

\begin{document}

\title{Wireless-OptSG: A Versatile Multi-Agent Optimization Framework for Wireless Resource Allocation}

\author{Wenhe~Zhang,~\IEEEmembership{Graduate Student Member,~IEEE}, Xin~Li,~\IEEEmembership{Member,~IEEE}, Hanzhi~Chen,~\IEEEmembership{Graduate Student Member,~IEEE}, and Chau~Yuen,~\IEEEmembership{Fellow,~IEEE}%
\thanks{Wenhe Zhang, Xin Li, Hanzhi Chen, and Chau Yuen are with the School of Electrical and Electronic Engineering, Nanyang Technological University, Singapore (e-mail: \{wenhe001, xin019, hanzhi001\}@e.ntu.edu.sg; chau.yuen@ntu.edu.sg).}}

\markboth{}{}

\maketitle

\begin{abstract}
6G wireless networks are expected to accommodate heterogeneous service requirements across diverse deployment scenarios, giving rise to physical-layer resource allocation problems with scenario-dependent objectives and constraints. Recent advances in large language models (LLMs) and agent systems can interpret wireless-domain specifications and incorporate execution feedback into optimization. However, translating language-level reasoning into feasible, high-quality resource allocations remains challenging, since candidates need to satisfy precise mathematical constraints throughout iterative search. This work proposes the Wireless Optimization Solution Generator (Wireless-OptSG), a versatile multi-agent framework for physical-layer resource allocation. Wireless-OptSG organizes optimization as a feedback-driven candidate search guided by a static meta-prompt and an evolving optimization history. In a nested outer-inner loop, parallel agents explore solution trajectories and periodically exchange high-quality records to guide candidate generation. Feasible initialization uses executable optimization primitives, while a tiered repair pipeline restores feasibility and solution-dependent context improves retained-record informativeness. Experiments on four representative tasks under perfect and imperfect channel state information show that Wireless-OptSG achieves performance comparable to scenario-specific state-of-the-art (SOTA) solvers, exceeds SOTA in some settings, and consistently outperforms generic baselines, demonstrating its potential as a transferable optimization framework for future wireless networks. 
\end{abstract}


\section{Introduction}
\IEEEPARstart{T}{he} evolution toward IMT-2030 is moving wireless networks from connectivity-oriented infrastructures to intelligent and service-adaptive systems. The ITU-R framework for IMT-2030 identifies  AI and communication, integrated sensing and communication (ISAC), and ubiquitous connectivity as representative usage scenarios, and further emphasizes ubiquitous intelligence as a major design trend for future mobile systems~\cite{iturM2160}. The corresponding technology-trend report also indicates that AI-based reasoning may support key components of future networks, including physical layer design and network architecture. Similar views have been reported in operator-oriented 6G statements and academic roadmaps~\cite{ngmn6gPosition}. These trends suggest that future wireless networks must support heterogeneous service requirements across diverse deployment scenarios, giving rise to a broad class of physical-layer resource allocation problems with scenario-dependent objectives and constraints.

Mathematical analysis and optimization have long served as fundamental tools for wireless resource allocation. These methods provide interpretable problem formulations, stable algorithmic behavior, and strong performance when the system model is well specified~\cite{liu2024optimizationSurvey}. They also allow feasibility and convergence to be examined through rigorous analysis. However, these advantages often rely on problem-specific reformulations and dedicated solver design. Developing an effective algorithm may require a long design cycle and substantial manual tuning, and the resulting solver is usually tied to the assumptions of a particular scenario. When the service objective, channel model, or constraint set changes, the original algorithm may require substantial redesign. Therefore, although mathematical optimization remains powerful, its dependence on handcrafted solver design limits its transferability across heterogeneous wireless scenarios.

To reduce the reliance on expert-designed algorithms, deep learning (DL) and reinforcement learning (RL) have been widely studied~\cite{letaief2022edgeAI}. DL methods typically learn a mapping from system parameters to optimization variables~\cite{shen2021gnnRRM,hu2022twoTimescale,schynol2023deepUnrolling,liang2024robustDL}, while generative models can learn conditional solution distributions from reference solutions~\cite{liang2025diffsg}. RL methods learn decision policies through reward-driven interactions and can reduce the need for labeled solutions~\cite{peng2021marlResource,tang2022drlOffloading}. Nevertheless, DL methods depend on the generation and quality of training data, while RL methods are sensitive to reward design and environment mismatch.

Recent progress in LLMs has motivated their use in wireless communications, where LLMs leverage broad prior knowledge and reasoning ability to interpret natural-language task descriptions and generate solutions. Existing studies increasingly treat LLMs as reasoning engines. Techniques such as chain-of-thought, ReAct, and Toolformer~\cite{wei2022chain,yao2023react,schick2023toolformer} enable LLMs and LLM-driven agents to decompose tasks and interact with external tools. Representative studies have applied these capabilities to multi-agent planning and automated wireless optimization~\cite{peng2025llmOptiRA,yang2026agenticprecoding,zhang2024genAIAgents,ni2026llmSpectrum,wei2026llmOptimizationSurvey,wen2026hybridRAG,guo2025llmAIGC,zhao2026edgeGI}. In this work, agent workflows that generate code, execute it, and use feedback to iteratively improve subsequent outputs are referred to as code-execution-feedback (CEF) methods. This paradigm reduces the need to manually derive a complete algorithm for each new problem, but several limitations still remain. Many existing studies mainly evaluate code executability or task completion, while systematic performance comparisons with scenario-specific SOTA solvers remain limited. Here, scenario-specific SOTA solvers denote problem-tailored mathematical optimization algorithms used as the strongest performance references. Moreover, wireless resource allocation problems with imperfect channel state information (CSI) often require problem-specific analytical treatments, such as inequality-based relaxation, conservative bound construction, or other transformations that convert the original formulation into a tractable optimization problem~\cite{yu2020robustIRS,sun2024risAntiJamming,liang2024robustDL}. These requirements cannot be fully addressed by simply selecting existing optimization algorithms.

These observations motivate a loop-engineered~\cite{osmani2026loopengineering} optimization framework. This framework should generalize across diverse wireless resource allocation problems under perfect and imperfect CSI, reduce reliance on task-specific training sets, and maintain solution quality comparable to SOTA solvers. Our main contributions are summarized as follows:

\begin{itemize}
    \item We propose Wireless-OptSG, a versatile multi-agent optimization framework for physical-layer wireless resource allocation. Rather than relying on task-specific post-training, manually designed scenario-specific optimizers, or direct LLM synthesis of a complete solver, Wireless-OptSG formulates the resource allocation problem as an iterative candidate search process guided by a static meta-prompt and an evolving optimization history. Its nested outer-inner loop supports parallel trajectory exploration, periodic exchange of high-quality records, and subsequent search refinement.

    \item We develop a set of modules for constructing feasible and informative solution records. The warm start initialization module builds a diverse initial exchange pool from executable optimization primitives, preventing the multi-agent search from starting with an empty or purely random history. The tiered repair pipeline maps extractable but infeasible LLM-generated candidates back to the feasible set with controlled deviation, thereby decoupling numerical feasibility restoration from LLM-based candidate proposal. The solution context generator further augments feasible records with optimization-relevant information to guide subsequent search.

    \item We evaluate the effectiveness of Wireless-OptSG on four representative physical-layer wireless resource allocation problems, covering both perfect CSI and imperfect CSI scenarios. Simulation results show that Wireless-OptSG achieves performance comparable to that of the corresponding scenario-specific SOTA solvers tailored to the same problems, outperforms conventional generic optimization baselines and CEF methods in the considered scenarios, and exceeds the SOTA reference at several tested settings in one case.
\end{itemize}

\section{System Model and Problem Formulation}\label{sec:system_model}
This section presents a system abstraction for wireless resource
allocation optimization problems. The following problem
formulations specify four benchmark problems considered
in this work. The detailed problem formulations and the corresponding SOTA solvers can be found in~\cite{salaun2020joint,kumar2023novel,mishra2022rate,choi2024joint}.

\subsection{System Model}

We consider a general wireless resource allocation problem that aims to determine controllable communication resources based on channel conditions and system parameters, so as to optimize a given performance metric under system constraints. The actual channel is denoted by $\mathbf H$, and the remaining non-channel parameters, such as noise powers and power budgets, are collected in $\boldsymbol\Omega$. The overall system state is denoted by $\Xi\triangleq(\mathbf H,\boldsymbol\Omega)$. For a given system state $\Xi$, let $\mathbf x$ denote the resource allocation variable, $\mathcal X(\Xi)$ denote the feasible set, and $f(\mathbf x;\Xi)$ denote the objective function. Since different problems may involve maximization or minimization, we use $\operatorname{opt}\in\{\max,\min\}$ to denote the optimization direction. The generic resource allocation problem is then written as
\begin{equation}
\begin{aligned}
\operatorname{opt}_{\mathbf x}\quad
&
f(\mathbf x;\Xi) \\
\mathrm{s.t.}\quad
&
\mathbf x\in\mathcal X(\Xi).
\end{aligned}
\label{eq:generic_deterministic_ra}
\end{equation}

For perfect CSI scenarios, \eqref{eq:generic_deterministic_ra} is directly applied. For imperfect CSI scenarios, the channel is modeled as $\mathbf H=\widehat{\mathbf H}+\widetilde{\mathbf H}$, where $\widehat{\mathbf H}$ is the estimated channel and $\widetilde{\mathbf H}$ is the channel estimation error. The estimated system state is denoted by $\widehat{\Xi}\triangleq(\widehat{\mathbf H},\boldsymbol\Omega)$. The conditional expected objective and the corresponding imperfect CSI formulation are
\begin{equation}
\begin{aligned}
\bar f(\mathbf x;\widehat{\Xi})
&=
\mathbb E_{\widetilde{\mathbf H}\mid\widehat{\mathbf H}}
\left[
f\bigl(\mathbf x;(\widehat{\mathbf H}+\widetilde{\mathbf H},\boldsymbol\Omega)\bigr)
\right], \\
\operatorname{opt}_{\mathbf x}\quad
&
\bar f(\mathbf x;\widehat{\Xi}) \\
\mathrm{s.t.}\quad
&
\mathbf x\in\mathcal X_{\mathrm{imp}}(\widehat{\Xi}),
\end{aligned}
\label{eq:imperfect_csi_ra}
\end{equation}
where $\mathcal X_{\mathrm{imp}}(\widehat{\Xi})$ denotes the feasible set.

\subsection{Representative Problem Instances}

To instantiate the general abstraction, we consider four representative problems. Unless otherwise specified, symbols are interpreted within the context of each problem. Throughout this subsection, $\mathbb C$, $\mathcal{CN}(\cdot,\cdot)$, and $\mathbf I_n$ denote the complex field, the circularly symmetric complex Gaussian distribution, and the $n\times n$ identity matrix, respectively.

\refstepcounter{subsubsection}
\addcontentsline{toc}{subsubsection}{\protect\numberline{\thesubsubsection}Problem 1: Joint Subcarrier and Power Allocation in Downlink NOMA}
\noindent\hspace{\parindent}{\normalfont\normalsize\itshape\thesubsubsectiondis\hskip 0.5em \textbf{Problem 1}: Joint Subcarrier and Power Allocation in Downlink NOMA}\par\nobreak
\label{subsubsec:problem_1}

We consider the non-orthogonal multiple access (NOMA) system in~\cite{salaun2020joint}, where a single base station (BS) serves $K$ users over $N$ orthogonal subcarriers. The user and subcarrier sets are denoted by $\mathcal K=\{1,\ldots,K\}$ and $\mathcal N=\{1,\ldots,N\}$. Subcarrier $n$ has bandwidth $W_n$, and the total bandwidth satisfies $\sum_{n\in\mathcal N}W_n=W$. Let $g_k^n$ and $\eta_k^n$ denote the channel gain and received noise power of user $k$ on subcarrier $n$. The channel gains are perfectly known, and the normalized noise power is defined as $\tilde{\eta}_k^n=\eta_k^n/g_k^n$.

The optimization variable is the transmit-power allocation
$\mathbf x=\mathbf p=(p_k^n)_{k\in\mathcal K,n\in\mathcal N}$, where $p_k^n\geq0$ denotes the power allocated to user $k$ on subcarrier $n$. The active-user set on subcarrier $n$ is denoted by
$\mathcal U_n\triangleq\{k\in\mathcal K\mid p_k^n>0\}$, and at most $M$ users can be multiplexed on each subcarrier. Users are decoded according to the successive interference cancellation (SIC) order $\pi_n$, which is determined by the normalized noise. Here, $\pi_n^{-1}(k)$ denotes the decoding position of user $k$ on subcarrier $n$. Under perfect SIC, the achievable rate of user $k$ on subcarrier $n$ is
\begin{equation}
R_k^n(\mathbf p^n)
=
W_n\log_2
\left(
1+
\frac{p_k^n}
{\sum_{j=\pi_n^{-1}(k)+1}^{K}p_{\pi_n(j)}^n+\tilde{\eta}_k^n}
\right),
\label{eq:noma_rate}
\end{equation}
where $\mathbf p^n=(p_k^n)_{k\in\mathcal K}$.

Let $w_k>0$ denote the priority weight of user $k$. The joint subcarrier and power allocation problem, which maximizes the weighted sum rate (WSR), can be formulated as
\begin{equation}
\begin{aligned}
\max_{\mathbf p}\quad
&
\sum_{k\in\mathcal K}
w_k
\sum_{n\in\mathcal N}
R_k^n(\mathbf p^n) \\
\mathrm{s.t.}\quad
&
\sum_{k\in\mathcal K}
\sum_{n\in\mathcal N}
p_k^n
\leq
P_{\max}, \\
&
\sum_{k\in\mathcal K}
p_k^n
\leq
P_{\max}^n,
\quad
\forall n\in\mathcal N, \\
&
p_k^n\geq 0,
\quad
\forall k\in\mathcal K,\ n\in\mathcal N, \\
&
|\mathcal U_n|\leq M,
\quad
\forall n\in\mathcal N.
\end{aligned}
\label{eq:noma_jspa}
\end{equation}

\vspace{\baselineskip}
\refstepcounter{subsubsection}
\addcontentsline{toc}{subsubsection}{\protect\numberline{\thesubsubsection}Problem 2: RIS-Assisted MU-MISO Downlink Transmit-Power Minimization}
\noindent\hspace{\parindent}{\normalfont\normalsize\itshape\thesubsubsectiondis\hskip 0.5em \textbf{Problem 2}: RIS-Assisted MU-MISO Downlink Transmit-Power Minimization}\par\nobreak
\label{subsubsec:problem_2}

We consider the problem in~\cite{kumar2023novel}, where an $N_t$-antenna BS serves $K$ single-antenna users in a multi-user multiple-input single-output (MU-MISO) downlink assisted by a reconfigurable intelligent surface (RIS) consisting of $N_s$ nearly passive reflecting elements. The user set and RIS-element set are denoted by $\mathcal K=\{1,\ldots,K\}$ and $\mathcal N_s=\{1,\ldots,N_s\}$, respectively. The BS-user, BS-RIS, and RIS-user channels associated with user $k$ are denoted by $\mathbf h_{t,k}\in\mathbb C^{1\times N_t}$, $\mathbf H_{t,s}\in\mathbb C^{N_s\times N_t}$, and $\mathbf h_{s,k}\in\mathbb C^{1\times N_s}$, respectively. The RIS reflection vector is denoted by $\boldsymbol\phi=[\phi_1,\ldots,\phi_{N_s}]^T\in\mathbb C^{N_s\times 1}$, where $\phi_{n_s}=\exp(-j\theta_{n_s})$, $\theta_{n_s}\in[0,2\pi)$, and $j$ is the imaginary unit. The corresponding reflection matrix is $\boldsymbol\Phi=\operatorname{diag}(\boldsymbol\phi)$.

Let $s_k$ denote the normalized information-bearing symbol intended for user $k$, with $\mathbb E[|s_k|^2]=1$, and let $\mathbf w_k\in\mathbb C^{N_t\times 1}$ denote the corresponding transmit beamforming vector. The transmit beamforming vectors are collected in $\mathbf w=[\mathbf w_1^T,\ldots,\mathbf w_K^T]^T$. The received signal at user $k$ is given by
\begin{equation}
y_k=
\left(
\mathbf h_{t,k}
+
\mathbf h_{s,k}\boldsymbol\Phi\mathbf H_{t,s}
\right)
\sum_{\ell\in\mathcal K}
\mathbf w_\ell s_\ell
+
\omega_k,
\end{equation}
where $\omega_k\sim\mathcal{CN}(0,\sigma_k^2)$ denotes circularly symmetric complex Gaussian noise. In this formulation, the users are assumed to have the same noise standard deviation $\sigma$, and the channels are normalized by $\sigma$. By reusing the same channel notation after normalization, the noise power in the signal-to-interference-plus-noise ratio (SINR) expression becomes one. The effective channel of user $k$ is then written as
\begin{equation}
\mathbf g_k
=
\mathbf h_{t,k}
+
\mathbf h_{s,k}\boldsymbol\Phi\mathbf H_{t,s},
\quad
\forall k\in\mathcal K.
\end{equation}
Under this normalized model, the SINR of user $k$ is
\begin{equation}
\gamma_k
=
\frac{
|\mathbf g_k\mathbf w_k|^2
}{
1+
\sum_{\ell\in\mathcal K\setminus\{k\}}
|\mathbf g_k\mathbf w_\ell|^2
}.
\end{equation}

The optimization variable is $\mathbf x=(\mathbf w,\boldsymbol\phi)$, which includes the active transmit beamforming vectors at the BS and the passive reflection coefficients at the RIS. Let $\Gamma_k$ denote the minimum required SINR of user $k$. The problem is formulated as
\begin{equation}
\begin{aligned}
\min_{\mathbf w,\boldsymbol\phi}\quad
&
\|\mathbf w\|_2^2 \\
\mathrm{s.t.}\quad
&
\gamma_k\geq \Gamma_k,
\quad
\forall k\in\mathcal K, \\
&
|\phi_{n_s}|=1,
\quad
\forall n_s\in\mathcal N_s.
\end{aligned}
\label{eq:irs_mumiso_powermin}
\end{equation}
where $\|\mathbf w\|_2^2=\sum_{k\in\mathcal K}\|\mathbf w_k\|_2^2$ denotes the total transmit power.

\vspace{\baselineskip}
\refstepcounter{subsubsection}
\addcontentsline{toc}{subsubsection}{\protect\numberline{\thesubsubsection}Problem 3: Precoder Optimization for RSMA Downlink MU-MIMO With Imperfect CSIT}
\noindent\hspace{\parindent}{\normalfont\normalsize\itshape\thesubsubsectiondis\hskip 0.5em \textbf{Problem 3}: Precoder Optimization for RSMA Downlink MU-MIMO With Imperfect CSIT}\par\nobreak
\label{subsubsec:problem_3}

We consider the rate-splitting multiple access (RSMA) problem in~\cite{mishra2022rate}, where an $M$-antenna BS serves $K$ users, each equipped with $Q$ receive antennas. The user set is denoted by $\mathcal K=\{1,\ldots,K\}$. Let $\mathbf H_k\in\mathbb C^{M\times Q}$ and $\widehat{\mathbf H}_k\in\mathbb C^{M\times Q}$ denote the actual and estimated channel matrices of user $k$. The actual channel is modeled as $\mathbf H_k=\widehat{\mathbf H}_k+\widetilde{\mathbf H}_k$, where $\widetilde{\mathbf H}_k$ is the error in the channel state information at the transmitter (CSIT). The aggregate actual and estimated channels are denoted by $\mathbf H=[\mathbf H_1,\ldots,\mathbf H_K]$ and $\widehat{\mathbf H}=[\widehat{\mathbf H}_1,\ldots,\widehat{\mathbf H}_K]$, respectively. The conditional distribution $f_{\mathbf H\mid\widehat{\mathbf H}}(\mathbf H\mid\widehat{\mathbf H})$ is assumed to be available at the BS. Here, the BS jointly encodes common and private streams and designs the aggregate precoder $\mathbf P=[\mathbf P_c,\mathbf P_1,\ldots,\mathbf P_K]$, where $\mathbf P_c\in\mathbb C^{M\times Q_c}$ is the common-stream precoder and $\mathbf P_k\in\mathbb C^{M\times Q_k}$ is the private-stream precoder for user $k$. The transmitted signal is $\mathbf x_{\mathrm{tx}}=\mathbf P\mathbf s$, where $\mathbb E[\mathbf s\mathbf s^H]=\mathbf I$, and the total transmit power satisfies $\operatorname{Tr}(\mathbf P\mathbf P^H)\leq P_t$.

For a given estimated channel $\widehat{\mathbf H}$, let $R_{c,k}^{\mathrm{avg}}(\widehat{\mathbf H};\mathbf P)$ and $R_{p,k}^{\mathrm{avg}}(\widehat{\mathbf H};\mathbf P)$ denote the conditional average common and private rates of user $k$, respectively, where the expectation is taken over $f_{\mathbf H\mid\widehat{\mathbf H}}(\mathbf H\mid\widehat{\mathbf H})$. The decodable common rate is
\begin{equation}
R_c^{\mathrm{avg}}(\widehat{\mathbf H};\mathbf P)
=
\min_{j\in\mathcal K}
R_{c,j}^{\mathrm{avg}}(\widehat{\mathbf H};\mathbf P).
\end{equation}

Let $\mathbf c=[C_1,\ldots,C_K]^T$ denote the common-rate allocation vector, where $C_k\geq0$ and $R_{k,\mathrm{tot}}^{\mathrm{avg}}=R_{p,k}^{\mathrm{avg}}(\widehat{\mathbf H};\mathbf P)+C_k$. Given user weights $\boldsymbol\mu=[\mu_1,\ldots,\mu_K]^T$, the weighted average sum-rate maximization (WASR) problem is formulated as
\begin{equation}
\begin{aligned}
\max_{\mathbf P,\mathbf c}\quad
&
\sum_{k\in\mathcal K}\mu_k R_{k,\mathrm{tot}}^{\mathrm{avg}} \\
\mathrm{s.t.}\quad
&
\sum_{k\in\mathcal K}C_k
\leq
R_c^{\mathrm{avg}}(\widehat{\mathbf H};\mathbf P), \\
&
\operatorname{Tr}(\mathbf P\mathbf P^H)\leq P_t, \\
&
C_k\geq0,
\quad
\forall k\in\mathcal K.
\end{aligned}
\label{eq:rsma_wasr}
\end{equation}
Here, the optimization variable is $\mathbf x=(\mathbf P,\mathbf c)$.

\vspace{\baselineskip}
\refstepcounter{subsubsection}
\addcontentsline{toc}{subsubsection}{\protect\numberline{\thesubsubsection}Problem 4: SCNR-Constrained MIMO-ISAC Beamforming With Imperfect CSIT}
\noindent\hspace{\parindent}{\normalfont\normalsize\itshape\thesubsubsectiondis\hskip 0.5em \textbf{Problem 4}: SCNR-Constrained MIMO-ISAC Beamforming With Imperfect CSIT}\par\nobreak
\label{subsubsec:problem_4}

We consider the problem in~\cite{choi2024joint}, where a BS equipped with $N$ antennas simultaneously serves $K$ single-antenna communication users and transmits dedicated radar waveforms for sensing. Let $\mathcal K=\{1,\ldots,K\}$ and $\mathcal M=\{1,\ldots,M\}$ denote the user set and radar-stream set. The joint beamforming matrix is $\mathbf F=[\mathbf F_C,\mathbf F_R]$, where $\mathbf F_C=[\mathbf f_{C,1},\ldots,\mathbf f_{C,K}]\in\mathbb C^{N\times K}$ contains the communication beamformers and $\mathbf F_R=[\mathbf f_{R,1},\ldots,\mathbf f_{R,M}]\in\mathbb C^{N\times M}$ contains the radar beamformers. The normalized transmit-power constraint is $\operatorname{Tr}(\mathbf F\mathbf F^H)\leq 1$.

The actual channel of user $k$ is modeled as
$\mathbf h_k=\widehat{\mathbf h}_k+\widetilde{\mathbf h}_k$, where
$\widehat{\mathbf h}_k\in\mathbb C^{N\times 1}$ is the estimated channel and
$\widetilde{\mathbf h}_k\sim\mathcal{CN}(\mathbf 0,\boldsymbol\Delta_k)$ is the CSIT error with covariance
$\boldsymbol\Delta_k=\mathbb E[\widetilde{\mathbf h}_k\widetilde{\mathbf h}_k^H]$. The aggregate estimated channel is denoted by
$\widehat{\mathbf H}=\{\widehat{\mathbf h}_k\}_{k\in\mathcal K}$.

For a given actual channel realization $\mathbf H=\{\mathbf h_k\}_{k\in\mathcal K}$ and beamforming matrix $\mathbf F$, the communication SINR and spectral efficiency (SE) of user $k$ are
\begin{equation}
\begin{aligned}
\gamma_k(\mathbf H;\mathbf F)
&=
\frac{
|\mathbf h_k^H\mathbf f_{C,k}|^2
}{
\sum_{i\in\mathcal K,i\neq k}
|\mathbf h_k^H\mathbf f_{C,i}|^2
+
\sum_{m\in\mathcal M}
|\mathbf h_k^H\mathbf f_{R,m}|^2
+
\sigma_k^2
}, \\
r_k(\mathbf H;\mathbf F)
&=
\log_2(1+\gamma_k(\mathbf H;\mathbf F)).
\end{aligned}
\label{eq:isac_se}
\end{equation}
Let $\mathbf G_{\mathrm{tar}}$ and $\mathbf G_{\mathrm{cl}}$ denote the target and clutter response matrices, respectively. The radar signal-to-clutter-plus-noise ratio (SCNR) is
\begin{equation}
R_{\mathrm{SCNR}}(\mathbf F)
=
\frac{
\operatorname{Tr}(\mathbf G_{\mathrm{tar}}\mathbf F\mathbf F^H\mathbf G_{\mathrm{tar}}^H)
}{
\operatorname{Tr}(\mathbf G_{\mathrm{cl}}\mathbf F\mathbf F^H\mathbf G_{\mathrm{cl}}^H)
+
N\sigma_R^2
}.
\label{eq:isac_scnr}
\end{equation}

The communication objective is the conditional average sum SE
$\bar{\Sigma}(\widehat{\mathbf H};\mathbf F)
=
\mathbb E_{\mathbf H\mid\widehat{\mathbf H}}
[\sum_{k\in\mathcal K}r_k(\mathbf H;\mathbf F)]$.
The problem is formulated as
\begin{equation}
\begin{aligned}
\max_{\mathbf F}\quad
&
\bar{\Sigma}(\widehat{\mathbf H};\mathbf F) \\
\mathrm{s.t.}\quad
&
R_{\mathrm{SCNR}}(\mathbf F)\geq \Gamma_{\mathrm{SCNR}}, \\
&
\operatorname{Tr}(\mathbf F\mathbf F^H)\leq 1.
\end{aligned}
\label{eq:isac_scnr_original}
\end{equation}
Here, the optimization variable is $\mathbf x=\mathbf F$.

\section{Proposed Wireless-OptSG Framework}\label{sec:method}

\subsection{Overview of Proposed Wireless-OptSG}
\label{subsec:framework}

\begin{figure}[H]
    \centering
    \includegraphics[width=0.98\linewidth, trim={7mm 88mm 6mm 8mm}, clip]{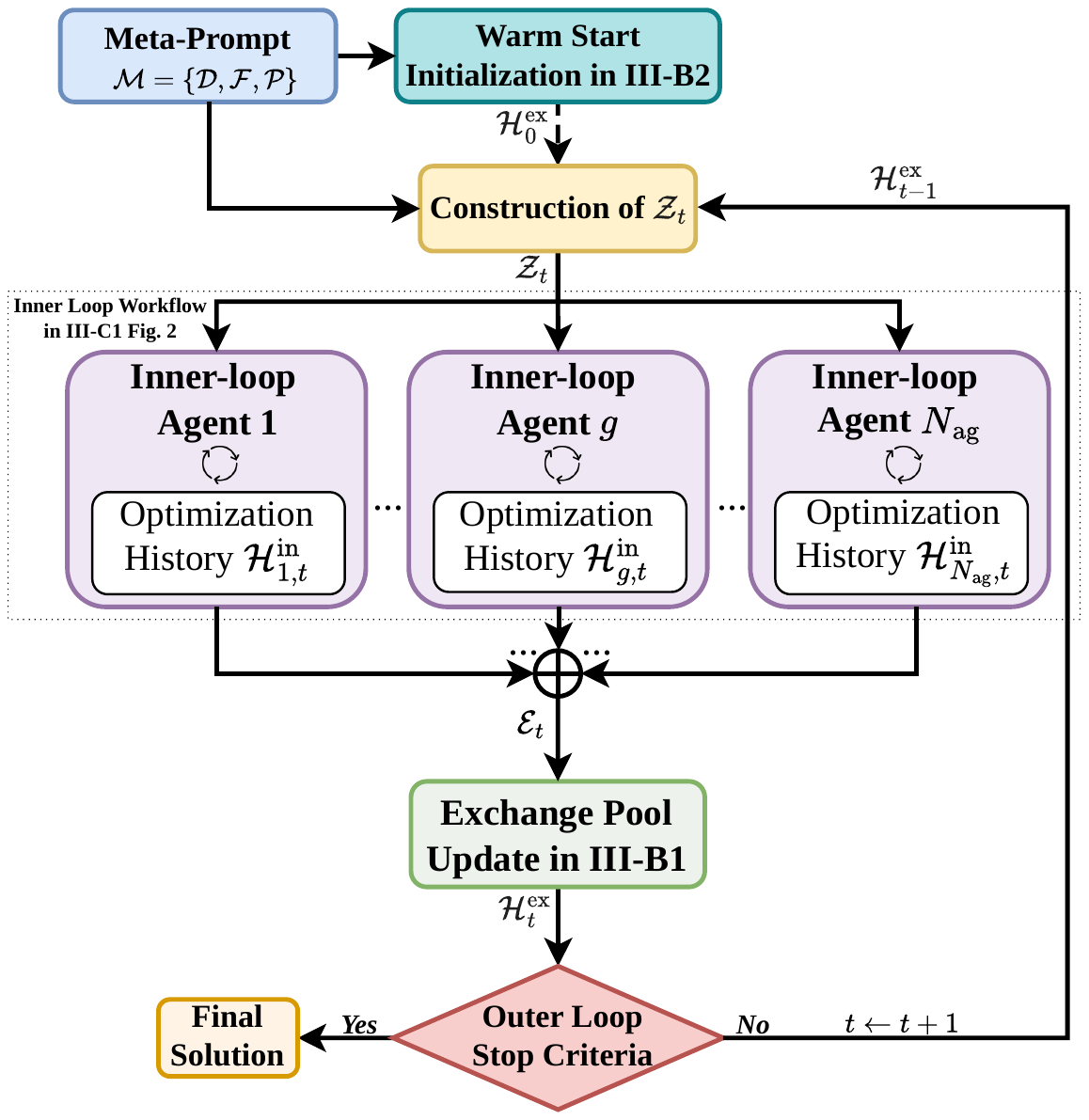}
\caption{Outer-loop workflow of Wireless-OptSG. The dashed connection carrying $\mathcal H_{0}^{\mathrm{ex}}$ from the warm start initialization module to the construction of $\mathcal Z_t$ is used only for initialization and is inactive in subsequent outer-loop iterations. $\oplus$ denotes the merge-and-deduplication operation described in Section~III-B1.}

    \label{fig:outer_loop}
\end{figure}

Wireless-OptSG is a multi-agent optimization framework with a nested-loop structure, as illustrated in Fig.~\ref{fig:outer_loop}. Following the terminology in~\cite{Yang2024LLMOptimizer}, the input is termed a meta-prompt; in this work, it contains only the static wireless resource allocation problem description and detailed system parameters, while the evolving optimization history is maintained separately. The output is the corresponding final solution. The outer loop coordinates multiple parallel agents and provides opportunities for the agents to exchange their optimization histories, while each agent internally performs an inner-loop candidate generation and evaluation process.

For compact notation, the meta-prompt is denoted by
\begin{equation}
\mathcal M
=
\{\mathcal D,\mathcal F,\mathcal P\},
\end{equation}
where $\mathcal D$ is the \textit{wireless communication scenario description}, which specifies the considered wireless resource allocation system; $\mathcal F$ is the \textit{objective function and constraints}, which specify the optimization direction, objective function, and corresponding constraint set; and $\mathcal P$ is the \textit{parameter definition}, which provides the numerical values, meanings, and dimensions of the variables required by the optimization problem. The optimization history that will be iteratively updated is a ranked set of candidate records, where each record contains a solution candidate, its objective value, and the associated solution context defined in Section~\ref{subsec:solution_context}.

\subsection{Outer Loop Workflow}

\subsubsection{\textbf{Exchange Pool Update}}

For the $t$\nobreakdash-th outer-loop iteration, where $t\in\{1,\ldots,T_{\mathrm{out}}\}$ and $T_{\mathrm{out}}$ is the maximum number of outer-loop iterations, a shared prompt is constructed for all parallel inner-loop agents as
\begin{equation}
\mathcal Z_t
=
\operatorname{Prompt}(\mathcal M,\mathcal H_{t-1}^{\mathrm{ex}}).
\end{equation}
Here, $\mathcal H_{t-1}^{\mathrm{ex}}$ denotes the optimization history retained in the exchange pool from the previous outer-loop iteration. The initial history $\mathcal H_0^{\mathrm{ex}}$ is constructed from $\mathcal M$ by the warm start initialization module. The use of $N_{\mathrm{ag}}$ parallel inner-loop agents is intended to mitigate the variability of LLM-based candidate generation and to enable cross-agent exploration. Even with the same input prompt, different LLMs or the same LLM may converge to different terminal candidates. By running multiple agents in parallel and merging their completed optimization histories after each outer-loop iteration, Wireless-OptSG allows high-quality candidates discovered along one trajectory to guide subsequent searches by other agents. Let $\mathcal H_{g,t}^{\mathrm{in}}$ denote the completed inner-loop optimization history produced by agent $g$ in the $t$\nobreakdash-th outer-loop iteration, where $g\in\{1,\ldots,N_{\mathrm{ag}}\}$. After all $N_{\mathrm{ag}}$ agents complete their inner-loop candidate-generation processes, the completed inner-loop optimization histories from the current $t$\nobreakdash-th outer-loop iteration are merged into the exchange pool:
\begin{equation}
\mathcal E_t
=
\operatorname{Dedup}
\left(
\bigcup_{g=1}^{N_{\mathrm{ag}}}
\mathcal H_{g,t}^{\mathrm{in}}
\right),
\end{equation}
where $\operatorname{Dedup}(\cdot)$ removes identical or numerically equivalent candidate records. The remaining candidates are sorted from worst to best according to their objective values under the specified optimization direction. The exchange pool optimization history is then updated by retaining the top-ranked solution candidates:
\begin{equation}
\mathcal H_t^{\mathrm{ex}}
=
\operatorname{Top}_{L_{\mathrm{ex}}}(\mathcal E_t),
\end{equation}
where $\operatorname{Top}_{L_{\mathrm{ex}}}(\cdot)$ returns the $L_{\mathrm{ex}}$ best solution candidate records. After the exchange pool is updated, Wireless-OptSG checks whether the outer-loop stopping conditions are satisfied. If neither condition is satisfied, $\mathcal H_t^{\mathrm{ex}}$ is combined with the static meta-prompt to construct $\mathcal Z_{t+1}$ for the next outer-loop iteration. The outer loop terminates when the number of completed iterations reaches $T_{\mathrm{out}}$ or the retained exchange-pool history is stable across three consecutive exchanges.

The average objective value of the exchange pool in the $t$-th outer-loop iteration is defined as
\begin{equation}
\bar{v}_t
=
\frac{1}{|\mathcal H_t^{\mathrm{ex}}|}
\sum_{i=1}^{|\mathcal H_t^{\mathrm{ex}}|}
v_{t,i},
\end{equation}
where $v_{t,i}$ is the objective value of the $i$\nobreakdash-th candidate record in $\mathcal H_t^{\mathrm{ex}}$, and $|\mathcal H_t^{\mathrm{ex}}|$ is the number of retained records. For $t\geq3$, the outer loop is regarded as converged if the relative change in $\bar{v}_t$ remains below the threshold $\epsilon_{\mathrm{out}}$ for two consecutive outer-loop transitions:
\begin{equation}
\frac{
\left|\bar{v}_{t}-\bar{v}_{t-1}\right|
}{
\left|\bar{v}_{t-1}\right|
}
\leq
\epsilon_{\mathrm{out}}
\quad
\text{and}
\quad
\frac{
\left|\bar{v}_{t-1}-\bar{v}_{t-2}\right|
}{
\left|\bar{v}_{t-2}\right|
}
\leq
\epsilon_{\mathrm{out}}.
\end{equation}
This condition is designed to reduce premature termination caused by a single small update. A small change between two adjacent exchange pools does not necessarily indicate that the outer-loop search has stabilized, because the newly retained records may still lead to a search for better candidates in later iterations. When either stopping condition is satisfied, the exchange pool returns the best-ranked candidate retained in the current exchange pool as the final solution.

\subsubsection{\textbf{Warm Start Initialization}}
\label{subsec:warm_start}

This module constructs $\mathcal H_0^{\mathrm{ex}}$ before the multi-agent search starts, providing initial candidate records that prevent the search from relying on an empty history or randomly generated solutions. For wireless optimization problems, mature standalone algorithms, such as regularized zero-forcing (RZF) and weighted minimum mean-square error (WMMSE), can serve as algorithmic primitives. The warm start initialization module transforms these primitives into multiple candidate-generation routines, thereby producing diverse initial references for the subsequent search.

The procedure is summarized in Algorithm~\ref{alg:warm_start}. Here, $N_{\mathrm{ws}}$ denotes the size of the high-quality temporary pool $\mathcal T_{\mathrm{top}}$, $T_{\mathrm{ws}}$ denotes the revision budget for each routine, and $\mathcal G$ denotes the feedback generated from execution errors or constraint violations. The warm start initialization module uses the meta-prompt $\mathcal M$ to identify applicable optimization primitives. Each routine may build on a single primitive or combine several primitives. Each routine is then refined through a CEF process, where $\mathcal G$ is used to revise the routine within the budget $T_{\mathrm{ws}}$. Only feasible candidates are evaluated and stored in the temporary pool $\mathcal T$.

After all routines are processed, the accepted candidates are merged and sorted according to their objective values under the specified optimization direction. The top $N_{\mathrm{ws}}$ candidates from $\mathcal T$ form a high-quality temporary pool $\mathcal T_{\mathrm{top}}$. However, directly using the $L_{\mathrm{ex}}$ best candidate records in $\mathcal T_{\mathrm{top}}$ may over-represent a single routine, since one routine can generate multiple candidates with similar structures. To balance candidate quality and diversity, Wireless-OptSG applies greedy farthest-point sampling to $\mathcal T_{\mathrm{top}}$, following the diversity-oriented subset selection idea in~\cite{shapira2024facecoresetnet}. The first selected candidate is the best candidate in $\mathcal T_{\mathrm{top}}$, and each subsequent candidate is selected as the one whose minimum distance to the already selected candidates is the largest. The selected candidates form the initial exchange pool optimization history:
\begin{equation}
\mathcal H_0^{\mathrm{ex}}
=
\operatorname{Select}_{L_{\mathrm{ex}}}(\mathcal T_{\mathrm{top}}),
\end{equation}
where $\operatorname{Select}_{L_{\mathrm{ex}}}(\cdot)$ denotes the above selection procedure.

\begin{algorithm}[H]
\caption{Warm Start Initialization}
\label{alg:warm_start}
\begin{algorithmic}[1]
\item[\textbf{Input:}] $\mathcal M$, $L_{\mathrm{ex}}$, $N_{\mathrm{ws}}$, $T_{\mathrm{ws}}$
\item[\textbf{Output:}] $\mathcal H_0^{\mathrm{ex}}$
\STATE Parse $\mathcal M$ and identify optimization primitives.
\STATE Construct multiple candidate-generation routines.
\STATE Initialize the temporary pool $\mathcal T\leftarrow\emptyset$.
\FOR{each routine}
\STATE Generate executable code and set $a\leftarrow0$.
\WHILE{$a<T_{\mathrm{ws}}$}
\STATE Execute the code to produce candidates.
\IF{feasible candidates are obtained}
\STATE Evaluate the feasible candidates.
\STATE Add the evaluated candidate records to $\mathcal T$.
\STATE \textbf{break}
\ELSE
\STATE $\mathcal G\leftarrow\operatorname{Feedback}(\text{errors},\text{violations})$.
\STATE $\text{code}\leftarrow\operatorname{Revise}(\text{code},\mathcal G)$; $a\leftarrow a+1$.
\ENDIF
\ENDWHILE
\ENDFOR
\STATE $\begin{aligned}
\mathcal T_{\mathrm{top}}
&=
\operatorname{Top}_{N_{\mathrm{ws}}}(\mathcal T).
\end{aligned}$
\STATE $\begin{aligned}
\mathcal H_0^{\mathrm{ex}}
&=
\operatorname{Select}_{L_{\mathrm{ex}}}(\mathcal T_{\mathrm{top}}).
\end{aligned}$
\RETURN $\mathcal H_0^{\mathrm{ex}}$.
\end{algorithmic}
\end{algorithm}

The warm start initialization module differs from the standard CEF workflow in its objective. A CEF process typically seeks the best-performing routine for a given problem setting. Consequently, routines that achieve strong but non-best performance are typically discarded. In contrast, the warm start initialization module retains feasible candidates produced by multiple routines, so that the initial optimization history can cover diverse solution structures and search directions. These retained candidates are not treated as final outputs. Instead, they serve as feasible initial references that guide the subsequent multi-agent search.

\subsection{Inner Loop Workflow}
\label{subsec:inner_loop_workflow}

\subsubsection{\textbf{Candidate Generation and Evaluation}}
\label{subsec:candidate_generation}

\begin{figure}[H]
    \centering
    \includegraphics[width=0.98\columnwidth,trim={16bp 195bp 17bp 28bp},clip]{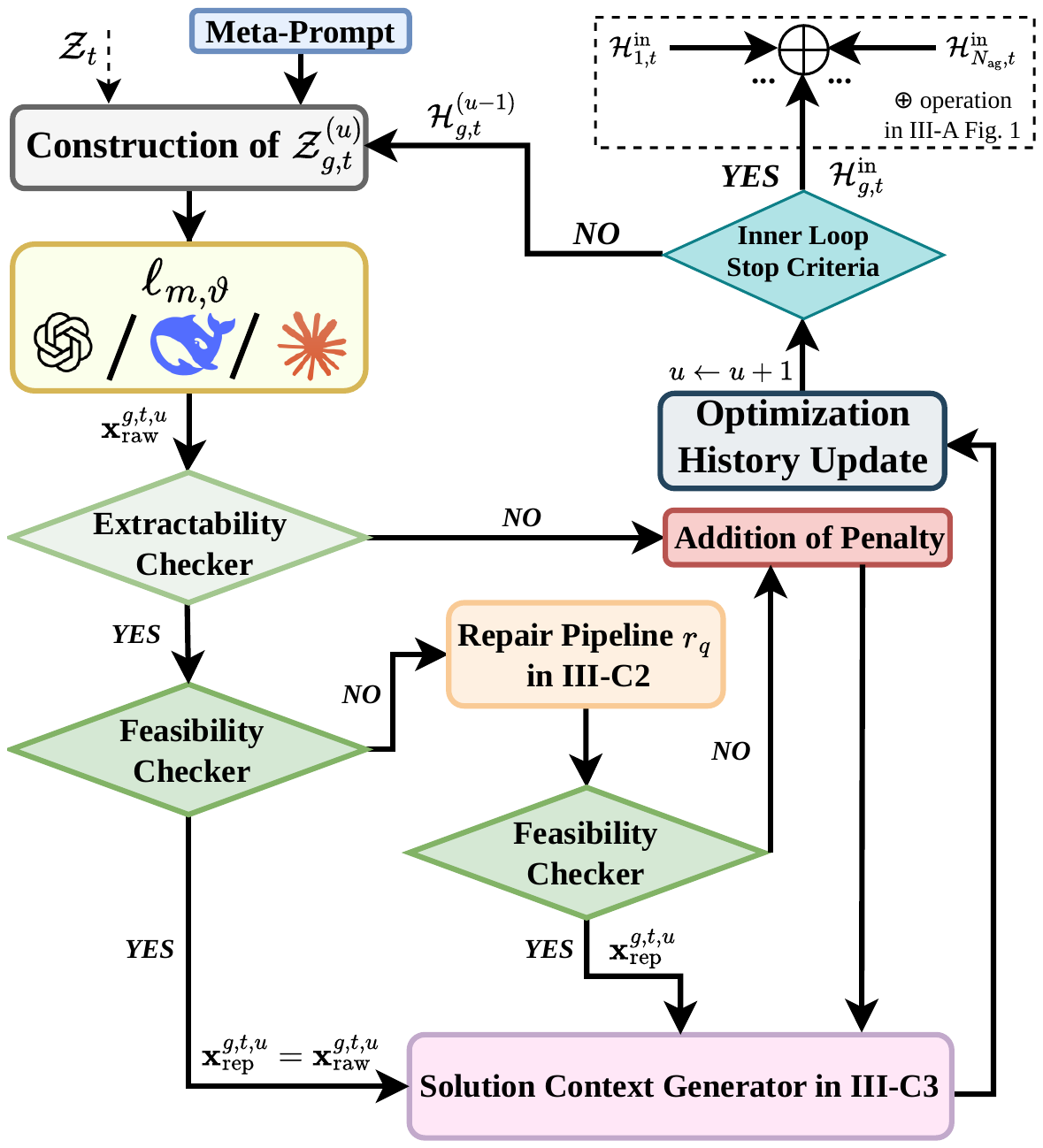}
    \caption{Workflow of an inner-loop agent for candidate generation, evaluation, and optimization history update in Wireless-OptSG. The dashed arrow from $\mathcal Z_t$ to the construction of $\mathcal Z_{g,t}^{(u)}$ indicates that the shared outer-loop prompt $\mathcal Z_t$ is supplied only once for initialization.}
    \label{fig:inner_loop}
\end{figure}

This part describes the procedure executed by each inner-loop agent. At the $t$\nobreakdash-th outer-loop iteration, all agents receive the complete prompt $\mathcal Z_t$. Each agent then performs an iterative candidate generation and evaluation process, as illustrated in Fig.~\ref{fig:inner_loop}. The goal of the inner loop is to use the candidate records in the current optimization history as references and generate new solution candidates with better objective values.

Each LLM-driven agent in the inner loop is specified by three components: an LLM selected from the available model pool, a temperature setting, and a repair pipeline defined in Section~\ref{subsec:repair_pipeline}. Let $\{\ell_1,\ldots,\ell_{N_{\mathrm L}}\}$ denote the set of candidate LLMs, $\{\vartheta_1,\ldots,\vartheta_{N_{\vartheta}}\}$ denote the set of temperature values, and $\{r_1,\ldots,r_{N_{\mathrm R}}\}$ denote the set of repair pipelines. In a full factorial deployment, these choices can form up to $N_{\mathrm L}N_{\vartheta}N_{\mathrm R}$ parallel agents. In practice, a subset can be selected according to the computational budget. The notation $\ell_{m,\vartheta}$ denotes the LLM $\ell_m$ operated under temperature $\vartheta$, and $r_q$ denotes the selected repair pipeline for the considered inner-loop agent.

The use of multiple temperature settings is motivated by the exploration--exploitation tradeoff observed in~\cite{Yang2024LLMOptimizer}. This study shows that the temperature parameter controls whether the optimizer LLM mainly adapts previously found solutions or explores candidates that differ more from the existing optimization history. Following this observation, Wireless-OptSG assigns different temperature settings to different inner-loop agents. Lower-temperature agents tend to exploit the current optimization history by producing candidates with smaller deviations from retained solutions, whereas higher-temperature agents encourage broader exploration by generating candidates with larger variations. This configuration reduces dependence on any single sampling trajectory by maintaining multiple complementary search paths over the same retained optimization history.

Consider agent $g$ in the $t$\nobreakdash-th outer-loop iteration. Let $\mathcal H_{g,t}^{(u)}$ denote the optimization history maintained by this agent after its $u$\nobreakdash-th inner-loop iteration, where $\mathcal H_{g,t}^{(0)}$ is initialized from $\mathcal H_{t-1}^{\mathrm{ex}}$. At the $u$-th inner-loop iteration, the agent-specific prompt is defined as $\mathcal Z_{g,t}^{(u)}\triangleq\operatorname{Prompt}(\mathcal M,\mathcal H_{g,t}^{(u-1)})$. Accordingly, $\mathcal Z_{g,t}^{(1)}=\mathcal Z_t$, while subsequent prompts incorporate the optimization history updated by the agent. The LLM then generates a raw output
\begin{equation}
\mathbf x_{\mathrm{raw}}^{g,t,u}
=
\ell_{m,\vartheta}
\left(\mathcal Z_{g,t}^{(u)}\right).
\end{equation}
Although the prompt requires the LLM to output a complete and feasible solution candidate, the raw output may be incomplete or violate the constraints. Each raw output first undergoes an extractability check, which verifies that all required variable blocks are present and can be parsed with the prescribed dimensions. For example, an output missing a required row or column is classified as nonextractable. A nonextractable output is assigned a sufficiently unfavorable score---a large negative value for a maximization problem or a large positive value for a minimization problem---and bypasses feasibility checking and repair. If $\mathbf x_{\mathrm{raw}}^{g,t,u}$ is extractable, it is checked by the feasibility checker. A candidate that passes this check is directly treated as the repaired candidate:
\begin{equation}
\mathbf x_{\mathrm{rep}}^{g,t,u}
=
\mathbf x_{\mathrm{raw}}^{g,t,u}.
\end{equation}
Otherwise, the candidate is sent to the selected repair pipeline $r_q$, which attempts to restore the feasibility of the raw output:
\begin{equation}
\mathbf x_{\mathrm{rep}}^{g,t,u}
=
r_q(\mathbf x_{\mathrm{raw}}^{g,t,u};\mathcal M).
\end{equation}
The repaired candidate is then checked again. If it remains infeasible after all three repair tiers, it is assigned the same unfavorable score as a nonextractable output.

Only an extractable and feasible candidate is evaluated under the scoring rule described below and converted into a candidate record. Let $v(\mathbf x_{\mathrm{rep}}^{g,t,u})$ denote its evaluated objective value. The candidate record $\mathcal R(\mathbf x_{\mathrm{rep}}^{g,t,u})$ contains the feasible candidate, its objective value, and the solution context. For such a candidate, the history update is written as
\begin{equation}
\mathcal H_{g,t}^{(u)}
=
\operatorname{Top}_{L_{\mathrm{in}}}
\left(
\mathcal H_{g,t}^{(u-1)}
\cup
\{\mathcal R(\mathbf x_{\mathrm{rep}}^{g,t,u})\}
\right),
\end{equation}
where $\operatorname{Top}_{L_{\mathrm{in}}}(\cdot)$ keeps the $L_{\mathrm{in}}$ best-ranked candidate records according to the specified optimization direction. During this update, nonextractable outputs and candidates that remain infeasible after repair are automatically excluded because their penalized scores are deliberately worse than those of feasible records. Operationally, the retained optimization history is unchanged for such an output, and the inner loop proceeds to the next iteration without interruption. Empirically, the inner-loop optimization history size is set to
\begin{equation}
L_{\mathrm{in}}
=
L_{\mathrm{ex}}+L_{\mathrm{buf}},
\end{equation}
where $L_{\mathrm{buf}}$ is a small buffer size. This buffer retains additional candidate records so that the average objective value of the inner-loop optimization history can still change when newly accepted candidates update the buffered records, which reduces the risk of prematurely satisfying the inner-loop stopping criteria. The inner loop terminates when either the maximum number of inner-loop iterations $T_{\mathrm{in}}$ is reached or the optimization history stops changing over multiple iterations. Specifically, let
\begin{equation}
\bar{v}_{g,t}^{(u)}
=
\frac{1}{|\mathcal H_{g,t}^{(u)}|}
\sum_{i=1}^{|\mathcal H_{g,t}^{(u)}|}
v_{g,t,i}^{(u)}
\end{equation}
denote the average objective value of the candidate records in $\mathcal H_{g,t}^{(u)}$, where $v_{g,t,i}^{(u)}$ is the objective value of the $i$\nobreakdash-th candidate record in this history. The inner loop is considered converged if the relative change in $\bar{v}_{g,t}^{(u)}$ remains no larger than a threshold $\epsilon_{\mathrm{in}}$ for $D_{\mathrm{in}}$ consecutive iterations:

\begin{equation}
\frac{
\left|
\bar{v}_{g,t}^{(u)}
-
\bar{v}_{g,t}^{(u-1)}
\right|
}{
\left|\bar{v}_{g,t}^{(u-1)}\right|
}
\leq
\epsilon_{\mathrm{in}},
\end{equation}
where multiple consecutive iterations are used because LLM outputs can vary even under the same model, temperature, and prompt. The additional patience gives the agent extra opportunities to generate candidates before the search terminates. When the inner loop of agent $g$ terminates at iteration $u$ in the $t$\nobreakdash-th outer-loop iteration, its completed optimization history is denoted by $\mathcal H_{g,t}^{\mathrm{in}}\triangleq\mathcal H_{g,t}^{(u)}$ and is subsequently merged into the exchange pool.

For perfect CSI problems, the evaluation of $\mathbf x_{\mathrm{rep}}$ is performed directly as
\begin{equation}
v(\mathbf x_{\mathrm{rep}})
=
f(\mathbf x_{\mathrm{rep}};\Xi).
\end{equation}

However, for imperfect-CSI problems, the exact conditional expectation is usually difficult to compute in closed form. Wireless-OptSG therefore adopts sample average approximation (SAA) to obtain the objective value. Given $B_{\mathrm{SAA}}$ conditional channel samples
$\{\mathbf H^{(b)}\}_{b=1}^{B_{\mathrm{SAA}}}$
drawn from $f_{\mathbf H\mid\widehat{\mathbf H}}(\mathbf H\mid\widehat{\mathbf H})$, with
$\Xi^{(b)}=(\mathbf H^{(b)},\boldsymbol\Omega)$, the SAA objective value is computed as
\begin{equation}
\widehat{f}_{B_{\mathrm{SAA}}}
(\mathbf x_{\mathrm{rep}};\widehat{\Xi})
=
\frac{1}{B_{\mathrm{SAA}}}
\sum_{b=1}^{B_{\mathrm{SAA}}}
f(\mathbf x_{\mathrm{rep}};\Xi^{(b)}).
\end{equation}

For the imperfect-CSI problems, SAA approximates the conditional expected objective through sampled channel realizations, thereby avoiding scenario-specific closed-form derivations during candidate evaluation and providing a unified scoring rule. Admittedly, SAA does not provide a worst-case robustness guarantee. It evaluates average performance under the assumed conditional channel distribution, and its accuracy depends on the number and quality of conditional samples. In this work, we assume that the channel-error distribution is known, as is commonly assumed by scenario-specific SOTA solvers, and that the number of samples is sufficiently large for the SAA estimate to approximate the conditional expected objective value.

\subsubsection{\textbf{Repair Pipeline Module}}
\label{subsec:repair_pipeline}

As described in Section~\ref{subsec:candidate_generation}, the extractability check precedes the repair stage, so the repair pipeline operates only on extractable raw candidates. Such a candidate $\mathbf x_{\mathrm{raw}}$ may contain meaningful solution patterns while still violating one or more constraints, since feasibility often depends on high-precision numerical relationships among many variables. An LLM may infer from the optimization history which variables should be adjusted but estimate the required adjustment magnitudes inaccurately. For example, a directionally reasonable update may slightly exceed the power budget. Wireless-OptSG therefore uses a repair pipeline to map $\mathbf x_{\mathrm{raw}}$ to a feasible candidate $\mathbf x_{\mathrm{rep}}$ whenever possible. This process neither replaces candidate generation nor solves the original problem directly; instead, it restores feasibility while preserving the structure and numerical values of $\mathbf x_{\mathrm{raw}}$ as much as possible.

To implement this objective, the repair pipeline adopts a three-tier hierarchy with increasing computational cost and intervention strength. \textit{Tier 1} contains low-complexity one-shot operations, including projection, rounding, normalization, and closed-form rescaling, for constraints that can be restored independently or through simple analytical corrections. \textit{Tier 2} contains lightweight iterative repairs for violations involving moderate variable coupling or nonconvex constraints. Representative operations include alternating projections, projected-gradient corrections, and trust-region updates around the raw candidate, with only a limited number of local correction steps. \textit{Tier 3} serves as an optimization-based fallback by constructing a feasibility-restoration problem that minimizes the deviation between $\mathbf x_{\mathrm{rep}}$ and $\mathbf x_{\mathrm{raw}}$ subject to the original constraints. For highly coupled or nonconvex constraints, this subproblem may use local approximations or convex relaxations implemented through techniques such as successive convex approximation (SCA) or semidefinite relaxation (SDR).

The three tiers are invoked sequentially according to a minimal-deviation principle. \textit{Tier 1} is attempted first because it is computationally inexpensive and typically introduces only small changes. \textit{Tier 2} is invoked only when \textit{Tier 1} is insufficient and limited local correction is required, while \textit{Tier 3} is reserved for violations involving strongly coupled variables or constraints. A candidate that remains infeasible after \textit{Tier 3} receives the unfavorable penalty score defined in Section~\ref{subsec:candidate_generation}. This ordering limits unnecessary modification of the LLM-generated search direction while progressively increasing the strength of feasibility restoration.

\begin{figure*}[!t]
    \centering
    \includegraphics[width=1\textwidth,trim={13bp 710bp 19bp 28bp},clip]{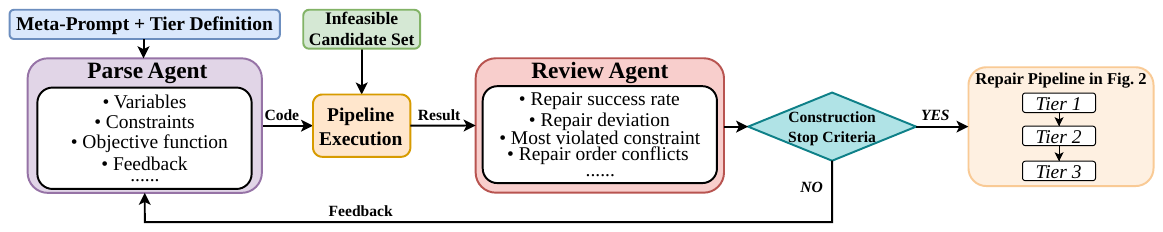}
    \caption{Workflow for the CEF-based construction and iterative refinement of tiered repair pipelines.}
    \label{fig:repair_pipeline}
\end{figure*}

The preceding description specifies how a repair pipeline operates during inner-loop candidate evaluation. Fig.~\ref{fig:repair_pipeline} instead illustrates how such pipelines are constructed through a CEF process. Given the meta-prompt and a tier definition specifying the available operator classes, their priority order, and the tier-upgrade conditions, Wireless-OptSG generates and iteratively refines executable repair pipeline code. The association between individual constraints and applicable repair algorithms is inferred during this process from the constraint structure specified in the meta-prompt. The corresponding infeasible candidate set is produced using the same candidate-generation mechanism as in the inner loop. After feasibility checking, extractable candidates that violate at least one constraint are retained as infeasible candidates. The repair pipeline code is evaluated on this set, and the execution results and repair diagnostics are used as feedback for subsequent revisions until the budget $T_{\mathrm{rp}}$ is reached.

After construction, Wireless-OptSG retains multiple repair pipelines rather than selecting a single dominant one, because the violation patterns of subsequent LLM-generated candidates are not known a priori. Different pipelines can therefore provide complementary repair capabilities for different infeasible candidates, increasing the likelihood that extractable candidates are restored to feasibility for subsequent evaluation. LLM-driven self-regeneration with feedback has also been explored in~\cite{noh2025adaptiveRAOLLMM,lee2026llmResourceAllocation}. However, repeated regeneration may still produce numerically infeasible candidates when variables are strongly coupled and constraints require precise numerical relationships; it may also lengthen the prompt as additional violation information is appended. By explicitly decoupling candidate generation from feasibility enforcement, Wireless-OptSG delegates numerical restoration to executable repair pipelines and reduces the burden on LLMs to resolve feasibility through repeated regeneration.

\subsubsection{\textbf{Solution Context Generator}}
\label{subsec:solution_context}

This module augments each feasible solution candidate with auxiliary quantities computed from the candidate and the available system information. For each $\mathbf x_{\mathrm{rep}}$, its solution context is denoted by
\begin{equation}
\boldsymbol{\psi}(\mathbf x_{\mathrm{rep}};\Xi)
=
\operatorname{Context}(\mathbf x_{\mathrm{rep}},\Xi),
\end{equation}
where $\operatorname{Context}(\cdot)$ is a deterministic post-processing operator that maps the candidate and the system state used for evaluation to a set of solution-dependent auxiliary quantities. These quantities are obtained by evaluating the objective expression, constraint functions, and relevant intermediate expressions specified in the meta-prompt, without introducing additional optimization variables. Examples include resource usage and constraint residuals. The candidate record stored in the optimization history is therefore written as
\begin{equation}
\mathcal R(\mathbf x_{\mathrm{rep}})
=
\left(
\mathbf x_{\mathrm{rep}},
v(\mathbf x_{\mathrm{rep}}),
\boldsymbol{\psi}(\mathbf x_{\mathrm{rep}};\Xi)
\right).
\end{equation}

The rationale for incorporating solution context is that a standard OPRO history in~\cite{Yang2024LLMOptimizer} typically records only candidate solutions and objective values, which do not reveal the candidates' proximity to constraint boundaries. This limitation is particularly important in wireless optimization because LLMs do not reliably infer high-precision auxiliary quantities from high-dimensional numerical candidates and may therefore propose apparently promising but infeasible solutions. For example, in rate maximization under a transmit-power budget, a high rate may result from power consumption near or beyond the allowable limit. Adding context entries such as total power consumption or per-block power allocation makes explicit how the objective value is achieved. The inner-loop agent can thus distinguish feasible improvements from candidates that benefit from constraint violations and direct subsequent search toward the remaining feasible margins.

\section{Performance Evaluation and Discussion}\label{sec:experiments}

\subsection{Simulation Setup}\label{subsec:experiment_setup}

\subsubsection{\textbf{Baselines}}\label{subsubsec:experiment_setup_baselines}

To compare Wireless-OptSG with representative generic methods, Table~\ref{tab:baseline_methods} reports the adopted baselines for the four considered problems. The first two baselines for each problem are mathematical methods applicable to the corresponding optimization formulation. The deep neural network (DNN) baseline follows the paradigm in~\cite{sun2018learning}. The training data are generated using the first two mathematical baselines, together with heuristic rules and closed-form formulas when applicable. A projection layer is applied to enforce the feasibility of the network outputs. The CEF baseline uses GPT-5.5 and follows the same construction process as the warm start initialization module. Because the CEF method is able to generate multiple optimization routines that can produce different feasible solutions under identical experimental settings, we choose to report the best objective value achieved among the routines to ensure a fair comparison.

\begin{table}[!htbp]
\centering
\begin{threeparttable}
\caption{Generic Baseline Solvers for Each Considered Problem}
\label{tab:baseline_methods}
\footnotesize
\begin{tabular*}{0.96\columnwidth}{@{\extracolsep{\fill}} c c c c c}
\hline
\textbf{Problem} & Baseline 1 & Baseline 2 & Baseline 3 & Baseline 4 \\
\hline
\textbf{1} & LDD & SCA & \multirow{4}{*}{DNN} & \multirow{4}{*}{CEF} \\
\textbf{2} & RCG & ADMM & & \\
\textbf{3} & WMMSE & RZF & & \\
\textbf{4} & SCA & SDR & & \\
\hline
\end{tabular*}
\begin{tablenotes}[flushleft]
\scriptsize
\item[] \textit{Note:} LDD, RCG, and ADMM denote Lagrangian dual decomposition, Riemannian conjugate gradient, and alternating direction method of multipliers, respectively.
\end{tablenotes}
\end{threeparttable}
\end{table}

\subsubsection{\textbf{Wireless-OptSG}}

In Wireless-OptSG, the candidate-generation model set $\{\ell_1,\ldots,\ell_{N_{\mathrm L}}\}$ is instantiated with DeepSeek-V4-Flash and GPT-5.6 Luna for the inner-loop search, whereas GPT-5.5 is used for the construction of warm start initialization module and repair pipeline module. For models that support batched generation, each call returns up to eight solution candidates. During all stages, no LLM or agent is permitted to perform web searches. All LLMs are accessed through application programming interface (API) calls. For each random channel realization, the complete nested outer-inner loop procedure of Wireless-OptSG is independently executed five times, and the best objective value is reported, following the same reporting protocol as the CEF method. The numerical values retained in the optimization history are recorded with eight decimal places. The remaining configuration parameters are summarized in Table~\ref{tab:wireless_optsg_config}. This setting yields $N_{\mathrm L}N_\vartheta N_{\mathrm R}=12$ parallel inner-loop agents.

\begin{table}[!htbp]
\caption{Main Configuration of Wireless-OptSG}
\label{tab:wireless_optsg_config}
\centering
\footnotesize
\setlength{\tabcolsep}{3pt}
\renewcommand{\arraystretch}{1.05}
\begin{tabular*}{0.92\columnwidth}{@{\extracolsep{\fill}}c c c c@{}}
\hline
\textbf{Symbol} & \textbf{Value} & \textbf{Symbol} & \textbf{Value} \\
\hline
$\{\vartheta_1,\ldots,\vartheta_{N_\vartheta}\}$ & $\{0.8,1.0,1.2\}$ & $N_{\mathrm R}$ & 2 \\
$T_{\mathrm{out}}$ & 30 & $T_{\mathrm{in}}$ & 100 \\
$L_{\mathrm{ex}}$ & 16 & $L_{\mathrm{in}}$ & 20 \\
$\epsilon_{\mathrm{out}}$ & 0.2\% & $\epsilon_{\mathrm{in}}$ & 0.5\% \\
$D_{\mathrm{in}}$ & 5 & $N_{\mathrm{ws}}$ & 80 \\
$T_{\mathrm{ws}}$ & 6 & $T_{\mathrm{rp}}$ & 6 \\
\hline
\end{tabular*}
\end{table}

\subsection{Case Study}\label{subsec:case_study}

\subsubsection{\textbf{Problem 1}}\label{subsubsec:case_study_problem_1}

In the simulations, the BS is located at the cell center, and users are randomly deployed in a hexagonal cell with diameter \(1000\)~m. The number of users is varied, while the number of subcarriers is fixed at \(N=10\). The total system bandwidth is \(5\)~MHz and is equally divided among the subcarriers. The carrier frequency is \(2\)~GHz, and the noise power spectral density is set to \(-174\)~dBm/Hz. The channel model accounts for distance-dependent path loss, log-normal shadowing with standard deviation \(10\)~dB, and Rayleigh fading. The user weights are independently generated from a uniform distribution over \([0,1]\). The total transmit power budget at the BS is \(P_{\max}=10\)~W, and the per-subcarrier power limit is set to \(P_{\max}^n=P_{\max}\) for all \(n\in\mathcal N\).

\begin{figure}[H]
\centering
\includegraphics[width=0.94\columnwidth,trim=7pt 8pt 7pt 7pt,clip]{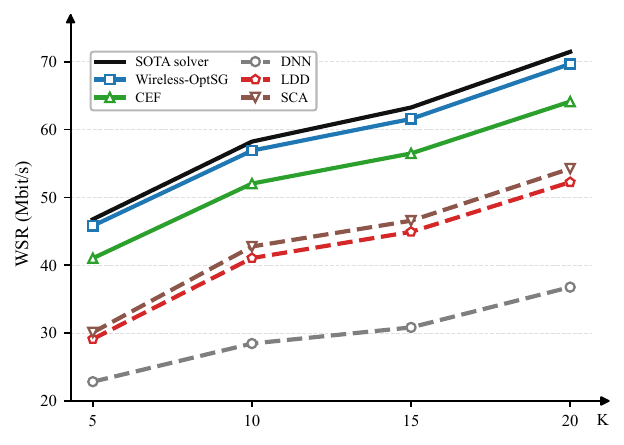}
\caption{WSR versus the number of users for Problem~1.}
\label{fig:noma_case1_wsr}
\end{figure}

Fig.~\ref{fig:noma_case1_wsr} reports the WSR achieved by different methods for Problem~1, with each point obtained by averaging over \(100\) random channel realizations for the corresponding number of users. As \(K\) increases from \(5\) to \(20\), Wireless-OptSG remains close to the SOTA solver and consistently outperforms the generic baselines. This gap between Wireless-OptSG and the SOTA solver does not noticeably widen as the number of users increases, which suggests that the performance of Wireless-OptSG is stable under the tested user loads. Among the generic baselines, CEF performs closest to Wireless-OptSG; the remaining gap reflects the benefit of subsequent multi-agent refinement beyond the standalone CEF-style search.

\subsubsection{\textbf{Problem 2}}\label{subsubsec:case_study_problem_2}

In this problem, a BS equipped with $N_t=4$ transmit antennas serves $K=4$ single-antenna users with the aid of a RIS. The minimum SINR requirement is set to $\Gamma_k=\Gamma=20$~dB for all users. The system operates at a carrier frequency of $2$~GHz with a total bandwidth of $20$~MHz, and the noise power spectral density is $-174$~dBm/Hz. The BS uses a uniform linear array (ULA) at $(0,20,10)$~m, while the RIS uses a uniform planar array (UPA) at $(30,0,5)$~m. The BS antennas and RIS reflecting elements are spaced by half a wavelength. The users are randomly distributed in a disk of radius $5$~m centered at $(350,10,2)$~m, with a minimum inter-user distance of twice the carrier wavelength. All BS-user, BS-RIS, and RIS-user channels are generated using Rician fading with distance-dependent path loss.

\begin{figure}[H]
\centering
\includegraphics[width=0.94\columnwidth,trim=7pt 9pt 7pt 7pt,clip]{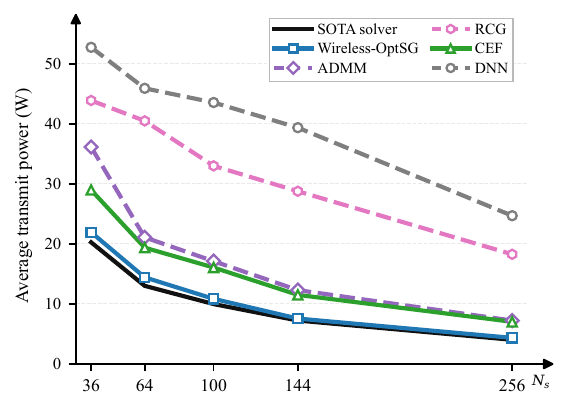}
\caption{Average transmit power versus the number of RIS elements for Problem~2.}
\label{fig:ris_case2_transmit_power}
\end{figure}

Fig.~\ref{fig:ris_case2_transmit_power} reports the average transmit power of different methods for Problem~2, with each point obtained by averaging over $100$ independent random channel realizations for the corresponding RIS size. The RIS is configured as a square UPA, and the number of reflecting elements is varied as $N_s=n_{\mathrm{row}}n_{\mathrm{col}}$, where $n_{\mathrm{row}}=n_{\mathrm{col}}\in\{6,8,10,12,16\}$, corresponding to $N_s\in\{36,64,100,144,256\}$. The results show that Wireless-OptSG remains close to the SOTA solver and consistently achieves lower transmit power than the generic baselines across all tested RIS sizes. These observations indicate that Wireless-OptSG can adapt to block-structured wireless optimization problems with coupled active and passive variables. They also suggest that the proposed search process can balance feasibility restoration and objective optimization, rather than merely satisfying the SINR constraints by increasing transmit power. Moreover, the consistent trend across $N_s$ supports the stability of Wireless-OptSG over the tested RIS dimensions.

\subsubsection{\textbf{Problem 3}}\label{subsubsec:case_study_problem_3}

For Problem~3, the considered system has a BS with $M=8$ transmit antennas serving $K=4$ users, each equipped with $Q=2$ receive antennas. The number of common streams is set to $Q_c=2$, and each user is assigned $Q_k=2$ private streams. Equal user weights are adopted in the WASR objective. The noise power is normalized as $\sigma_n^2=1$, and the signal-to-noise ratio (SNR) is set to $15~\mathrm{dB}$. The BS has imperfect CSIT, whereas perfect channel state information is assumed at the receivers. The entries of \(\widetilde{\mathbf H}_k\) are modeled as independent and identically distributed random variables following \(\mathcal{CN}(0,\sigma_e^2)\). The channel variance is normalized as $\sigma_k^2=1$, and the CSIT quality is parameterized by the exponent $\alpha$, with the CSIT-error variance given by $\sigma_e^2=P_t^{-\alpha}$, which means a larger $\alpha$ leads to more accurate CSIT.

For each $\alpha$, the reported value is obtained using a two-stage Monte Carlo evaluation protocol. We first generate $100$ independent estimated channel realizations $\widehat{\mathbf H}$. For each estimated channel realization, $1000$ conditional channel samples are drawn from $f_{\mathbf H\mid\widehat{\mathbf H}}(\mathbf H\mid\widehat{\mathbf H})$ and used for SAA-based scoring. After convergence, the best candidate in the final exchange pool according to its SAA score is evaluated on an independent set of $100$ true channel realizations drawn from the same conditional distribution. The resulting objective values are averaged first over these true channel realizations and then over the $100$ estimated channel realizations.

\begin{figure}[H]
\centering
\includegraphics[width=0.98\columnwidth,trim=7pt 9pt 7pt 7pt,clip]{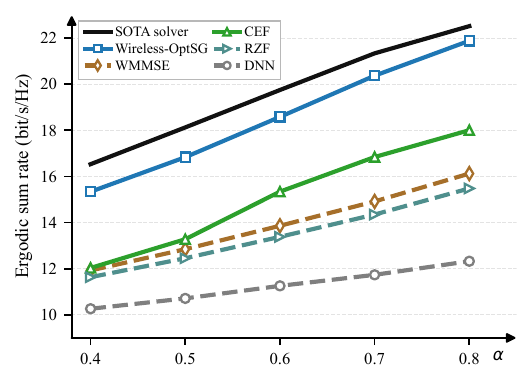}
\caption{Ergodic sum rate versus $\alpha$ for Problem~3.}
\label{fig:rsma_case3_wsr}
\end{figure}

Fig.~\ref{fig:rsma_case3_wsr} shows that Wireless-OptSG achieves the strongest performance among the generic baselines and remains closest to the SOTA solver across the tested range of $\alpha$. The performance gap between Wireless-OptSG and the SOTA solver varies modestly across the tested settings. These variations may be partly attributed to finite-sample variability in SAA-based scoring. Because $\alpha$ controls the CSIT-error variance, a fixed finite sample set may represent the corresponding conditional channel distribution with different accuracy across CSIT-quality levels. Nevertheless, SAA-based scoring remains useful for guiding the search under the considered imperfect-CSIT setting.

\subsubsection{\textbf{Problem 4}}\label{subsubsec:case_study_problem_4}

The ISAC BS is equipped with $N=8$ antennas and serves $K=4$ single-antenna communication users with $M=8$ dedicated radar waveform streams. The communication and radar noise powers are normalized as $\sigma_c^2=\sigma_R^2=1$, and the transmit SNR is set to $25$~dB. We set the SCNR threshold to $\Gamma_{\mathrm{SCNR}}=14$~dB. The communication channel covariance is generated from a ULA-based spatially correlated multipath model. For each user, the covariance matrix $\mathbf K_k$ is constructed from $L=5$ paths, where the mean angle of departure is uniformly drawn from $[-\pi/3,\pi/3]$ and the angular spread is $10^\circ$. The radar scene contains one target at $\theta_{\mathrm{tar}}=\pi/6$ and three clutter sources at $\{-\pi/3,-\pi/8,\pi/3\}$, with unit reflection coefficients. To instantiate the imperfect CSIT model, we generate $\widehat{\mathbf h}_k\sim\mathcal{CN}(\mathbf 0,(1-\kappa^2)\mathbf K_k)$ and $\widetilde{\mathbf h}_k\sim\mathcal{CN}(\mathbf 0,\kappa^2\mathbf K_k)$, yielding the true channel $\mathbf h_k=\widehat{\mathbf h}_k+\widetilde{\mathbf h}_k$ and the CSIT-error covariance $\boldsymbol{\Delta}_k=\kappa^2\mathbf K_k$. Thus, a larger $\kappa$ corresponds to less accurate CSIT.

\begin{figure}[H]
\centering
\includegraphics[width=0.98\columnwidth,trim=7pt 9pt 7pt 7pt,clip]{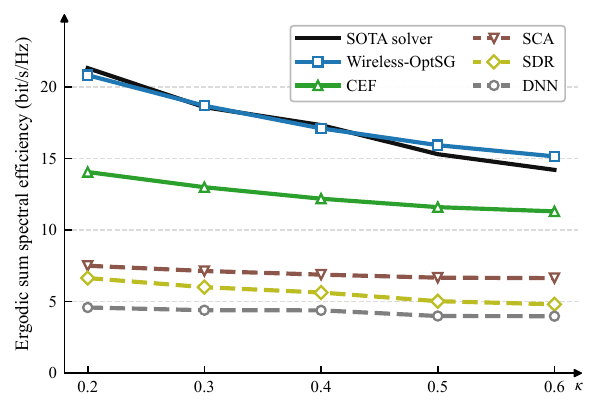}
\caption{Ergodic sum spectral efficiency versus $\kappa$ for Problem~4.}
\label{fig:isac_case4_esse_vs_kappa}
\end{figure}

Fig.~\ref{fig:isac_case4_esse_vs_kappa} reports the ergodic sum spectral efficiency under different CSIT error levels. As in Section~\ref{subsubsec:case_study_problem_3}, for each value of $\kappa$, we first generate $100$ estimated channel realizations. For each estimated channel, $1000$ conditional channel samples are drawn from the corresponding error distribution for SAA-based scoring. After Wireless-OptSG converges for the given estimated channel, another $100$ true channel realizations are sampled from the same conditional distribution for evaluation. The objective values obtained for these true channels are averaged to obtain the result for this estimated channel, and the final plotted point is obtained by averaging over the $100$ estimated channel realizations.

Under this setting, Wireless-OptSG consistently outperforms the generic baselines across the $\kappa$ values and remains closest to the SOTA solver. Notably, it slightly exceeds the adopted SOTA reference at several tested points. This observation may be attributed to two factors. First, iterative mathematical solvers for nonconvex formulations are often tied to initialization-dependent stationary trajectories, which may limit their ability to move between different local regions. Second, the adopted SOTA solver relies on a transformed surrogate problem, where the scaling-based lower-bound approximation may introduce conservatism relative to the original objective. For Wireless-OptSG, candidates are evaluated using an SAA-based surrogate, and the search starts from diverse feasible candidates generated by different routines. Together with the exchange mechanism and patience-based stopping rules, this design provides opportunities to refine candidates from different local regions, which may allow Wireless-OptSG to find feasible solutions with higher objective values than the adopted SOTA reference in some cases.

\subsection{Ablation Study}\label{subsec:ablation_study}

\begin{figure*}[!t]
\centering
\includegraphics[width=0.81\textwidth,trim=6pt 7pt 6pt 7pt,clip]{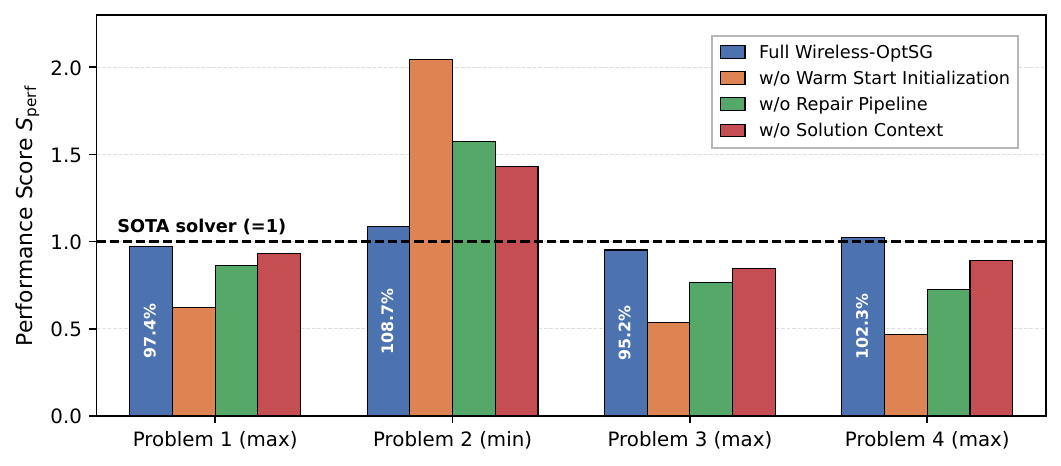}
\caption{Ablation performance scores for the four considered problems. Max and Min denote maximization and minimization, respectively; higher scores are better for Max, whereas lower scores are better for Min. Percentages on the full Wireless-OptSG bars show normalized performance relative to SOTA.}
\label{fig:ablation_performance_score}
\end{figure*}

\begin{table*}[!t]
\centering
\caption{Average Number of Outer-Loop Iterations at Termination for Different Ablation Variants}
\label{tab:ablation_outer_iterations}
\footnotesize
\renewcommand{\arraystretch}{1.15}
\begin{tabular}{c c c c c}
\hline
\textbf{Problem}
& \textbf{Full Wireless-OptSG}
& \textbf{w/o Warm Start}
& \textbf{w/o Repair Pipeline}
& \textbf{w/o Solution Context} \\
\hline
\textbf{Problem~1} & 20.96 & 13.74 & 3.00 & 17.42 \\
\textbf{Problem~2} & 25.28 & 18.13 & 3.00 & 23.05 \\
\textbf{Problem~3} & 21.82 & 10.95 & 3.00 & 19.71 \\
\textbf{Problem~4} & 27.63 & 15.86 & 3.00 & 23.09 \\
\hline
\end{tabular}
\end{table*}

To isolate the contribution of each key component, we conduct an ablation study by comparing the full Wireless-OptSG with three variants, namely \emph{w/o warm start}, \emph{w/o repair pipeline}, and \emph{w/o solution context}. The ablation study follows the same simulation setting as the corresponding case study in Section~IV-B. The performance score is computed through a two-stage averaging procedure. Specifically, for each system configuration, the objective value achieved by a given method under each channel realization is first normalized by the objective value achieved by the corresponding SOTA solver under the same channel realization. These normalized ratios are then averaged over all channel realizations under this configuration to obtain an intermediate score. Finally, the intermediate scores obtained under all tested configurations are averaged to obtain the final performance score:
\begin{equation}\label{eq:ablation_intermediate_score}
\bar{S}_{s}
=
\frac{1}{|\mathcal{H}_s|}
\sum_{h\in\mathcal{H}_s}
\frac{v_{s,h}^{\mathrm{method}}}{v_{s,h}^{\mathrm{SOTA}}},
\end{equation}
\begin{equation}\label{eq:ablation_performance_score}
S_{\mathrm{perf}}
=
\frac{1}{|\mathcal{S}|}
\sum_{s\in\mathcal{S}}
\bar{S}_{s}.
\end{equation}
Here, $\mathcal{S}$ denotes the set of tested system configurations for the considered problem, and $s\in\mathcal{S}$ denotes one specific configuration. The set $\mathcal{H}_s$ contains all channel realizations generated under configuration $s$, and $h\in\mathcal{H}_s$ denotes one channel realization. The terms $v_{s,h}^{\mathrm{method}}$ and $v_{s,h}^{\mathrm{SOTA}}$ denote the objective values obtained by the considered method and the SOTA solver under the same configuration $s$ and channel realization $h$, respectively. The quantity $\bar{S}_{s}$ is the intermediate normalized score under configuration $s$, and $S_{\mathrm{perf}}$ is the final performance score after averaging over all tested configurations.

As shown in Fig.~\ref{fig:ablation_performance_score}, the full Wireless-OptSG achieves the best performance score among all variants for each problem. Its normalized performance remains within 5\% and 8\% of the SOTA references for the maximization and minimization problems, respectively. Problem~2 is a transmit-power minimization problem with relatively small SOTA objective values; hence, its ratio-based score should be interpreted together with the absolute transmit-power results in Fig.~\ref{fig:ris_case2_transmit_power}.

Table~\ref{tab:ablation_outer_iterations} reports the average number of outer-loop iterations at termination. The averaging follows the same two-stage procedure as the performance score. For each fixed problem setting, we first record the number of completed outer-loop iterations for each channel realization and average these values over all channel realizations. The resulting setting-wise averages are then averaged over all tested settings of the corresponding problem. This metric complements the performance score by reflecting how many outer-loop exchanges are needed before the search terminates.

For the full Wireless-OptSG, the average number of iterations at termination provides additional insight into the search difficulty across problems. Problems with high-dimensional variables or strongly coupled constraints may require more exchanges before the retained exchange pool stabilizes. However, this quantity should not be interpreted as a standalone measure of search quality, since it is also affected by problem structure, channel realizations, and the quality of initialized candidates and repair pipelines. In particular, fewer outer-loop iterations do not necessarily imply more effective convergence, because an ablated variant may terminate early when few useful candidates are accepted and the exchange pool stagnates. Therefore, Table~\ref{tab:ablation_outer_iterations} should be interpreted together with Fig.~\ref{fig:ablation_performance_score}: the performance score reflects final solution quality, while the iteration count helps contextualize whether termination corresponds to productive convergence or early stagnation. The module-wise analysis further examines this point.

\subsubsection{\textbf{w/o Warm Start Initialization}}

We first examine the effect of removing the warm start initialization module. Without this module, the initial exchange pool contains no candidate records generated from executable optimization routines. As a result, the early-stage search largely resembles zero-shot candidate generation, where the LLM must infer both the solution structure and numerical values from the textual problem description and the specified output format. This setting can be less reliable for problems with high-dimensional variables, since format instructions alone may not ensure complete variable blocks or consistent output structures. Although the repair pipeline can still restore feasibility for raw candidates with compatible dimensions, such candidates may provide weak optimization guidance if their underlying solution structures are not sufficiently informative.

The performance degradation of the w/o warm-start variant supports the role of this module in providing a high-quality and diverse initial exchange pool. By collecting feasible candidates produced by different optimization routines, the warm start initialization module offers multiple initial solution structures that may correspond to different stationary regions or, more generally, different local regions of the search space. In contrast, when this module is removed, the inner-loop search can still build an optimization history over later iterations, but this history is initialized from weaker and less diverse candidates. This also helps explain why the w/o warm-start variant can terminate in fewer outer-loop iterations in Table~\ref{tab:ablation_outer_iterations} while still achieving inferior performance scores in Fig.~\ref{fig:ablation_performance_score}: the exchange pool may stabilize earlier around lower-quality records. Therefore, Wireless-OptSG does not discard mature optimization primitives; instead, it uses them as informative starting points and further improves upon them through the subsequent multi-agent search process.

\subsubsection{\textbf{w/o Repair Pipeline}}

We next examine the effect of removing the repair pipeline. Although Fig.~\ref{fig:ablation_performance_score} shows that the w/o repair pipeline variant does not always incur the largest performance degradation, Table~\ref{tab:ablation_outer_iterations} reveals a distinct behavior: for all considered problems, the average number of outer-loop iterations at termination reaches $3.00$, which is the minimum value allowed by the stopping rule. This suggests that effective refinement after warm start initialization is strongly limited without the repair pipeline. Although the inner-loop agents can still generate raw outputs from the optimization history generated by the warm start initialization module, most extractable candidates fail the feasibility check. As a result, too few new feasible records are inserted into the history, and the exchange pool stabilizes early.

\begin{table}[!htbp]
\centering
\caption{Feasibility Outcomes for the Full Wireless-OptSG}
\label{tab:candidate_repair_outcomes}
\scriptsize
\renewcommand{\arraystretch}{1.12}
\begin{tabular*}{0.96\columnwidth}{@{\extracolsep{\fill}}c c c c@{}}
\hline
\textbf{Problem} & \textbf{Directly Feasible} & \textbf{Feasible After Repair} & \textbf{Penalized Outputs} \\
\hline
\textbf{Problem~1} & 12.14\% & 76.34\% & 11.52\% \\
\textbf{Problem~2} & 22.78\% & 67.71\% & 9.51\% \\
\textbf{Problem~3} & 16.91\% & 72.46\% & 10.63\% \\
\textbf{Problem~4} & 6.39\% & 82.71\% & 10.90\% \\
\hline
\end{tabular*}
\end{table}

This interpretation is further supported by Table~\ref{tab:candidate_repair_outcomes}, which summarizes the outcomes of raw LLM outputs $\mathbf x_{\mathrm{raw}}^{g,t,u}$. Since the tier used for successful repair depends on the constraint structure of each problem and on the selected repair pipeline, we focus on aggregate repair outcomes rather than tier-level proportions. Here, ``Directly Feasible'' denotes extractable candidates that already satisfy all constraints, ``Feasible After Repair'' denotes initially infeasible candidates that are restored to feasibility by the repair pipeline, and ``Penalized Outputs'' denotes outputs that either remain infeasible after all repair tiers or cannot be parsed into complete variable sets. Across the considered problems, only a limited fraction of raw outputs are directly feasible, whereas a substantial fraction can be made feasible through repair. Without the repair pipeline, early stabilization mainly reflects insufficient candidate utilization rather than effective convergence.

\subsubsection{\textbf{w/o Solution Context Generator}}

We finally examine the effect of removing the solution context generator. As shown in Fig.~\ref{fig:ablation_performance_score}, the w/o solution context variant causes the smallest performance degradation among the three ablated variants. This is consistent with the fact that the warm start initialization module and repair pipelines are still retained, so the framework can still generate feasible candidates and update the optimization history. Nevertheless, the remaining gap between this variant and the full Wireless-OptSG indicates that objective values alone provide less informative guidance than candidate records augmented with solution-dependent context. Table~\ref{tab:ablation_outer_iterations} further shows that the w/o solution context variant terminates in fewer outer-loop iterations than the full Wireless-OptSG, suggesting that the exchange pool may stop improving earlier when the retained history lacks auxiliary information. Overall, the solution context generator improves the informativeness of the optimization history and helps sustain effective refinement, leading to better final performance.

\subsection{Current Limitations and Future Directions}

Although Wireless-OptSG demonstrates its generality and effectiveness on the considered problems, several limitations remain. First, scalability is limited by the context needed to represent high-dimensional, high-precision candidates, as in large-scale antenna systems. If the history can hold only one complete candidate, the LLM could refine one variable block at a time, following the block-wise search idea in~\cite{romera2024mathematical}, while external code accepts only feasible, objective-improving updates. A simpler alternative is to await models with longer context windows. Second, Wireless-OptSG does not directly solve worst-case robust optimization or min--max formulations, which generally require problem-specific bounds, adversarial channel construction, or tractable reformulations. Since no universal procedure can derive tight bounds for arbitrary problems, human expertise remains necessary. Wireless-OptSG can be applied after reformulation but does not replace this analytical step. Third, the computational time remains non-negligible because the workflow requires repeated candidate generation, feasibility checking, repair, and outer-loop exchange, despite parallel agents and relatively small LLMs. Wireless-OptSG is therefore better suited to offline optimization than to strict real-time control. Its offline outputs could further support scenario-dependent databases or lightweight surrogate models for low-latency online use.

\section{Conclusion}
\label{sec:conclusion}

This paper presented Wireless-OptSG, a versatile multi-agent optimization framework for physical-layer wireless resource allocation. Following the loop-engineering principle, Wireless-OptSG formulates optimization as a feedback-driven candidate search, in which parallel agents explore solution trajectories within a nested outer-inner loop and periodically exchange high-quality records. Feasible warm-start records, tiered repair pipelines, and solution context jointly improve search initialization, candidate utilization, and history informativeness. Evaluations on four representative problems under perfect and imperfect CSI show that Wireless-OptSG remains within 5\% and 9\% of the adopted SOTA references for the maximization and minimization problems, respectively, exceeds the SOTA reference in one maximization case, and consistently outperforms the considered generic optimization and CEF baselines. These results demonstrate that Wireless-OptSG can reduce reliance on task-specific training data and handcrafted solver design while producing feasible, high-quality solutions across heterogeneous wireless resource allocation scenarios.

\end{document}